\documentclass[10pt,journal,final]{IEEEtran}
\usepackage{mathrsfs}
\usepackage{bbding}
\usepackage{graphicx}
\usepackage{float}
\usepackage{booktabs}
\usepackage[hidelinks]{hyperref}
\usepackage{cite}

\usepackage{enumitem}
\usepackage{color}
\usepackage{ulem}
\usepackage{psfrag}
\usepackage{subfigure}
\usepackage{amssymb}
\usepackage{amsthm}
\usepackage{setspace}
\usepackage{epsfig}
\usepackage{pifont}
\usepackage{amsmath}
\usepackage{array}
\newcommand{\PreserveBackslash}[1]{\let\temp=\\#1\let\\=\temp}
\newcolumntype{C}[1]{>{\PreserveBackslash\centering}p{#1}}
\newcolumntype{R}[1]{>{\PreserveBackslash\raggedleft}p{#1}}
\newcolumntype{L}[1]{>{\PreserveBackslash\raggedright}p{#1}}
\usepackage{multicol}
\usepackage{multirow}
\usepackage{diagbox}
\usepackage{pifont}
\usepackage{indentfirst}
\usepackage{amsfonts}
\usepackage{algorithm}
\usepackage{algorithmicx}
\usepackage{algpseudocode}
\floatname{algorithm}{Procedure}
\usepackage{fancyhdr}
\usepackage{amscd}
\usepackage[cmyk]{xcolor}
\usepackage{bm}
\usepackage{fancyhdr}
\setlist{itemsep=0pt,parsep=0pt}

\allowdisplaybreaks[4]

\IEEEoverridecommandlockouts

\usepackage{caption}
\begin{document}
	\title{\huge Diffusion Model for Multiple Antenna Communications}
	
	\author{
		% \thanks{This work is supported by National Natural Science Foundation of China (NSFC) under Grant XXX}
		% \IEEEauthorblockN{}.
		\thanks{Jia Guo and Xiaoxia Xu is with the School of Electronic Engineering and Computer
			Science, Queen Mary University of London, London E1 4NS, U.K. (e-mail:
			jia.guo@qmul.ac.uk, x.xiaoxia@qmul.ac.uk).
			
			Yuanwei Liu is with the Department of Electrical and Electronic Engineering,
			The University of Hong Kong, Hong Kong (e-mail: yuanwei@hku.hk).
			%			
			%			Hyundong Shin is with the Department of Electronic Engineering, Kyung Hee University, Seoul, South Korea (e-mail:
			%			hshin@khu.ac.kr).
			%	
			
			Arumugam Nallanathan is with the School of Electronic Engineering and Computer
			Science, Queen Mary University of London, London E1 4NS, U.K. (e-mail:
			a.nallanathan@qmul.ac.uk).}
		%			
		%			A part of this work has been accepted by IEEE Globecom 2024 Workshops \cite{guo2024multi}. If this paper is accepted, we will publicize our codes on GitHub.}
	%		%	
	\IEEEauthorblockN{Jia Guo, \emph{Member, IEEE}, Xiaoxia Xu, \emph{Member, IEEE}, Yuanwei Liu, \emph{Fellow, IEEE}, \\and Arumugam Nallanathan, \emph{Fellow, IEEE}}
	
	%\IEEEauthorblockA{Queen Mary University of London\\ \{jia.guo, yuanwei.liu, a.nallanathan\}@qmul.ac.uk} 
}
\maketitle
\setcounter{page}{1}
\thispagestyle{empty}

\begin{abstract}
	The potential of applying diffusion models (DMs) for multiple antenna communications is discussed. A unified framework of applying DM for multiple antenna tasks is first proposed.  Then, the tasks are innovatively divided into two categories, i.e., decision-making tasks and generation tasks, depending on whether an optimization of system parameters is involved. For each category, it is conceived 1) how the framework can be used for each task and 2) why the DM is superior to traditional artificial intelligence (TAI) and conventional optimization tasks. It is highlighted that the DMs are well-suited for scenarios with strong interference and noise, excelling in modeling complex data distribution and exploring better actions. A case study of learning beamforming with a DM is then provided, to demonstrate the superiority of the DMs with simulation results. Finally, the applications of DM for emerging multiple antenna technologies and promising research directions are discussed.
	
\end{abstract}

\section{Introduction}\label{sec:intro}
The advancement of wireless communications has introduced cutting-edge techniques, including simultaneously transmitting and reflecting reconfigurable intelligent surface (STAR-RIS) and continuous aperture array (CAPA). By leveraging higher frequencies, wider bandwidths, and denser antenna arrays, these techniques help support higher network throughput, enabling emerging applications such as the metaverse, digital twins, and e-health across far-field, near-field, and hybrid near- and far-field scenarios \cite{NFC-survey}. Meanwhile, 
these advancements significantly increase the complexity of wireless systems, highlighting the need for intelligent and adaptive solutions.

Wireless artificial intelligence (AI) has become a prominent technique in sixth-generation (6G) communications, with deep learning serving as the major tool. Due to the low inference complexity for real-time implementation and the ability of joint optimization with channel acquisition, deep neural networks (DNNs) have been widely applied in wireless communications. Traditional artificial intelligence (TAI) primarily focuses on learning deterministic policies, where actions are uniquely determined by given states. However, in the next-generation wireless communication networks characterized by advanced techniques, diverse applications and novel scenarios, the randomness in wireless systems has been greatly increased. To be more specific,
\begin{itemize}
	\item High-frequency band utilization increases susceptibility to environmental factors such as scattering, diffraction, and absorption, resulting in frequent channel variations.
	\item The deployment of large antenna arrays presents challenges for accurate channel estimation.
	\item The diversity of applications involves various types of edge devices with complex noise and strong inter-device interference.
\end{itemize}
These randomnesses make many policies such as beamforming and resource allocation non-deterministic, posing challenges for TAI approaches.

Recently, a new class of deep learning architectures, called generative foundation model (GFM), has been proposed and used for generating new content that mirrors the training data. Popular GFMs include generative adversarial networks (GANs), where a generator produces synthetic samples to compete with a discriminator. Another widely used GFM is variational encoder (VAE), which encodes data into latent space and decodes it back to generate realistic outputs. Compared to TAI models, GFMs show great promise in addressing non-determinism by learning complex prior data distributions, which can serve as a priori knowledge for improved decision-making \cite{Survey-DM-DRL-NetwOpt-2024,GvsD}.

Among GFMs, diffusion models (DMs) have recently emerged as a powerful tool capable of generating high-fidelity data through a step-by-step stochastic denoising process, accurately approximating complex data distributions \cite{DM-DRL}. Recent works have applied DMs for wireless tasks including resource management \cite{Survey-DM-DRL-NetwOpt-2024, XXX}, signal detection \cite{GAI-signal-detection-TCOM2021} and direction of arrival (DoA) estimation for integrated sensing and communication \cite{GAI-ISAC-WCM2024}. Nonetheless, the understanding of the applicability of DMs in wireless communications, especially multiple antenna communications, is still in the early stages. A unified framework of training DMs for wireless tasks is also yet to be developed. To harness the potential of DMs and chart a path forward, we discuss why and how DMs can be applied to multiple antenna communications and emerging technologies. More specifically, \emph{1) to answer the question of why}, we reveal that DMs excel at \emph{modeling} complex data distributions and \emph{exploring} better actions, which are suitable for learning non-deterministic policies characterized by strong noise and interference.
\emph{2) To answer the question of how}, we propose a unified framework for training DMs in different methods. 
We further outline the promising applications of DM-aided framework in emerging multiple antenna technologies. We also highlight the research directions of applying DMs for multiple antenna communications.

\section{Framework of DM}\label{sec:DM-framework}
\begin{figure*}
	\centering
	\includegraphics[width=.9\linewidth]{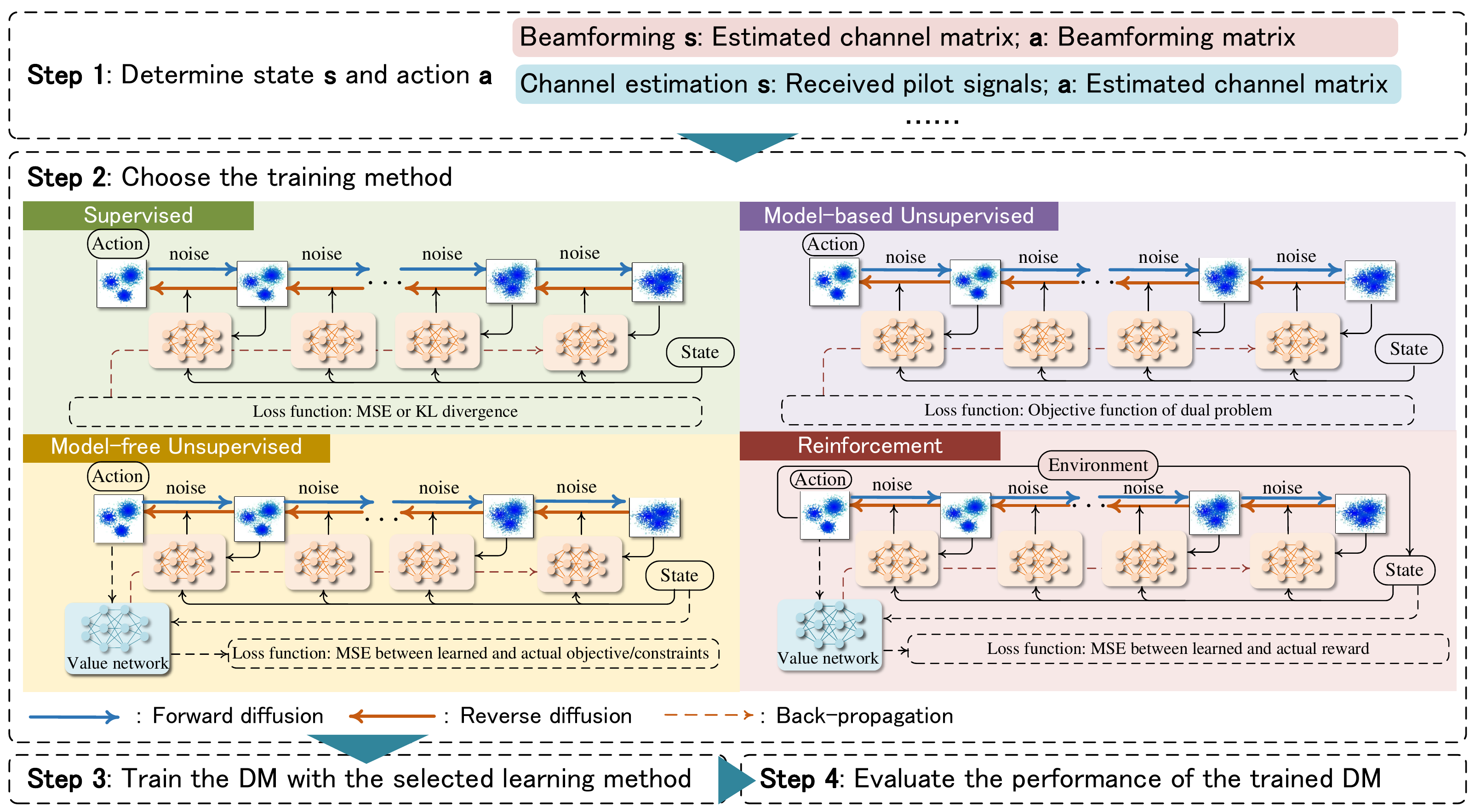}
	\caption{Framework of learning policies with DMs.}
	\label{fig:framework}
\end{figure*}
This section provides the principle of the diffusion model (DM) and introduces a comprehensive framework for policy learning with DMs.

A DM typically consists of two key processes, a forward diffusion process of adding noise to the original data, followed by a reverse diffusion process for denoising, as illustrated in each sub-figure of training method in Fig. \ref{fig:framework} (say the sub-figure of supervised learning).
\begin{itemize}
	\item \textbf{Forward diffusion process}: An original data  (say a image) undergoes multiple steps, where a Gaussian noise is added in each step. The final output approximates pure Gaussian noise.
	\item \textbf{Reverse diffusion process}: A DNN is trained to denoise the final output, by predicting the noise added in each step during the forward diffusion process. Specifically, in each step, the mean value and standard deviation of the added noise are learned by the DNN (which is referred to as noise-prediction DNN). Then, a Gaussian noise with the learned mean value and standard deviation is added on each step to generate the output of the previous step. Such a process allows for the generation of new data instead of only recovering the original data.
\end{itemize}

Through step-by-step noise prediction, DMs capture the complex distributions of the original data, facilitating the generation of high-fidelity new data. By incorporating a condition into the denoising DNN, DMs can generate data conditioned on specific inputs.

In essence, DMs model the probability of the generated data given the conditions. By interpreting generated data as actions and conditions as states, the DM is analogous to a TAI model (say DNN) for policy learning, where a policy is the mapping from states to actions. 
% However, DMs offer unique advantages by (i) modeling intricate action distributions and (ii) exploring and generating new actions following these distributions.

\subsection{Training Methods of DMs}

DMs can serve as an alternative module of TAI models, which can be trained in supervised, model-based/model-free unsupervised and reinforcement paradigms. In what follows, we outline methods to train DMs for learning policies.

\subsubsection{Supervised Learning}
This is how DMs are trained when it is used for image generation tasks \cite{DDPM}. This learning method requires the labeled data, i.e., the expected actions, which is also the original data at the start of the forward diffusion process. After adding noise at each step in the forward diffusion process, the denoising DNN can be trained to minimize the mean-squared-error (MSE) between the added and predicted noise \cite{Survey-DM-DRL-NetwOpt-2024}, or the Kullback-Leibler (KL) divergence between their respective distributions \cite{DDPM}. 

\subsubsection{Model-based Unsupervised Learning}
When labels are unavailable or only sub-optimal, unsupervised learning can be used to train the DMs. For example, for learning a multi-user power allocation policy in an interference network to maximize spectral efficiency (SE) with maximal power budget, the state is the estimated channel matrix and the action is the allocated power. If optimal actions are unknown, the DMs can be trained to maximize the SE over all the training samples, and the learned power can be passed through a normalization layer to satisfy the power constraint. 

For more complex constraints in a problem that cannot be satisfied simply by normalization, the Lagrange multiplier method can be leveraged, where the original optimization problem is firstly transformed into a dual problem without constraints of original variables. Then, the objective function of the dual problem averaged over all the training samples is used as the loss function for training the DM \cite{Eisen_Unspv_TSP2019}.

The DM is trained with multiple epochs.
In each epoch, a better action is firstly updated with gradient descent. Then, the better action is used as the original data to generate the noise-polluted data  in each forward diffusion step. Finally, the DM is trained, again to predict the noise.

\subsubsection{Model-Free Unsupervised Learning}
When the expressions of the objective functions and constraints of an optimization problem are not available, the loss function for unsupervised learning cannot be expressed. In this case, a \emph{value network} can be trained in a supervised manner to approximate the objective function and constraints of a problem \cite{SCJ-modelfree}. After the value network is trained, the gradients via the loss function can be obtained via back-propagation for training the DM. The DM and the value network can be alternatively trained to optimize a policy that can maximize the reward.

\subsubsection{Reinforcement Learning}
Some learning tasks can be modeled as a Markov decision process with multiple Markov steps, which are distinct from the ``diffusion steps'' in the forward and reverse diffusion processes. The action taken in each Markov step affects the state in the next Markov step. Deep reinforcement learning methods such as the deep Q-network (DQN) can be used to learn the actions of multiple steps. The DM-based DQN is with a nested architecture with multiple Markov steps, and each Markov step has multiple diffusion steps. Similar to the model-free unsupervised learning, a value network (also called Q-network in the literature) needs to learn the reward of each Markov step.

\subsection{General Framework of Applying DMs}
The general framework of applying DMs for multiple antenna communications can be summarized in the following steps, as illustrated in Fig. \ref{fig:framework}.
\begin{enumerate}
	\item For a given task in multiple antenna communications, the state and action are firstly recognized, and the DM aims to learn the policy, i.e., the conditional probability of actions given states.
	\item Then, the learning method is selected based on the characteristics of the learning task and the availability of the labels and the expressions of objective function and constraints, if an optimization problem can be formulated for the task. Specifically, if the task can be regarded as a Markov decision process, then deep reinforcement learning should be used. Otherwise, if the labels can be obtained, then supervised learning can be used. If the labels are not available, regarding whether the expressions of objective function and constraints are available, model-based and model-free unsupervised learning can respectively be used.
	\item The DM is trained with the selected learning method.
	\item The performance of the trained DM is evaluated in the inference phase.
\end{enumerate}

\subsection{Categorization of Tasks in Multiple Antenna Communications}

To address the diverse tasks in multiple antenna communications systematically, we propose a criterion to divide the tasks in multiple antenna communications into two categories, generation and decision-making tasks, depending on whether system parameters need to be optimized. To be more specific, generative tasks focus on estimating or reconstructing information from received signals, while decision-making tasks focus on optimizing system parameters based on the estimated data. 

In what follows, we introduce the applications of DMs in the two types of tasks. We emphasize that: 
\begin{itemize}
	\item For generation tasks, DMs are superior to TAI and conventional optimization methods in modeling data distribution, which helps better reconstruct data from noise-polluted or insufficient measurements.
	\item For decision-making tasks, DMs can not only model data distribution but also explore better actions to find more satisfactory solutions to non-convex problems in interference scenarios.
\end{itemize}
Hence, \emph{DMs are well-suited for scenarios with strong noise and interference}. For instance, in satellite communication systems, the receiving signals are often weak and overwhelmed by noise, while in heterogeneous networks with massive BSs and users, interference levels are significantly high.

\section{Generation Tasks}

This section focuses on the DM-empowered generation for multiple antenna communications. 
There are two fundamental generation tasks in multiple antenna communications, namely channel estimation 
and signal detection.

\vspace{-1mm}
\subsection{Channel Estimation}

Channel estimation generally recovers unknown channel parameters 
by observing the pilot symbols that are transmitted to probe channel effects and environment. 
DMs can be trained in a \textit{supervised manner} to estimate and predict multi-antenna channels.

Conventional channel estimation methods may simplify the non-ideal multi-antenna channel effects 
caused by multi-path fading, time-spatial non-stationarity, and 
pilot contamination, which are challenging to model in practice. 
However, this leads to non-determinism in channel estimation.
With the expressive representation ability, DMs can \textit{accurately model the underlying distribution of realistic multi-antenna channels through the iterative denoising process}. 
Then, the modeled distribution serves as a priori knowledge to generate high-fidelity channel matrix from limited pilots.  
More specifically, DMs empower high-accuracy and overhead-reduced channel estimation for multiple antenna communications in several ways: 
\begin{itemize}
	\item \textbf{Overcoming multi-path fading and contamination:}
	DMs can iteratively denoise accurate channels even when the 
	received pilot signals suffer from intricate multi-path propagation and pilot contamination.
	By capturing the interactions between different propagation paths and the effects of non-orthogonal pilots, 
	more accurate channel estimates can be achieved. 
	\item \textbf{Handling sparse pilot measurements:}
	Instead of requiring dense pilot symbols across all antennas, 
	DMs can generate fine-grained estimates from limited pilot symbols by learning meaningful spatial structures of multi-antenna channel matrices, which helps reduce estimation overhead. 
	\item \textbf{Approximating complex realistic channel distributions:} 
	DMs can represent mixed-distribution multi-antenna channels in time-varying real-world scenes (e.g., indoor-to-outdoor, air-to-ground) 
	by modeling channels' statistical properties without relying on predefined static/single scenarios.
	By training DMs on large ray-tracing datasets, 
	the divergence between the simulated distribution and the complex time-varying distribution of realistic channels can be minimized.
\end{itemize}

\vspace{-1mm}

\subsection{Signal Detection}
Signal detection aims to recover desired signals from noisy signals observed at receivers, 
which can be tackled by DMs leveraging \textit{supervised learning} methods. 
Multiple antenna techniques enhance signal recovery reliability and spectral efficiency by spatial diversity and spatial multiplexing. 
However, this requires processing signals transmitted by different antennas and separating the overlapping signals at receivers. 
Hence, it suffers from non-determinism and nonlinear disturbances incurred by large-scale and small-scale fading, multi-antenna interference, and potential channel estimation errors. 
DMs can improve signal detection in the following aspects: 
\begin{itemize}
	\item \textbf{Denoising polluted signals}: DMs can overcome aforementioned highly nonlinear disturbances 
	through its unique denoising nature, thus refining noisy signals into cleaner and structured data. 
	This can enhance detection accuracy even in low signal-to-noise ratio (SNR) regimes, 
	where conventional signal detection methods may fail to separate useful signals. 
	\item \textbf{Learning signal distributions}:
	The signal detection becomes an ill-conditioned problem in highly spatial-correlated channels (e.g., in urban hotspots with strong interference). 
	In this case, the corrupted signals correspond to multiple potential solutions, rendering the inefficiency of conventional detection techniques.
	The probabilistic denoising architecture allows DMs to learn inherent signal structures, 
	thus distinguishing the desired signal even if the problem is ill-conditioned.
\end{itemize}

\vspace{-1mm}
\section{Decision-Making Tasks}\label{sec:decision-tasks}
In this section, we demonstrate the application of diffusion models (DMs) to decision-making tasks in multiple antenna communications, including beamforming, beam training, and resource allocation. 
\begin{figure*}
	\centering
	\includegraphics[width=\linewidth]{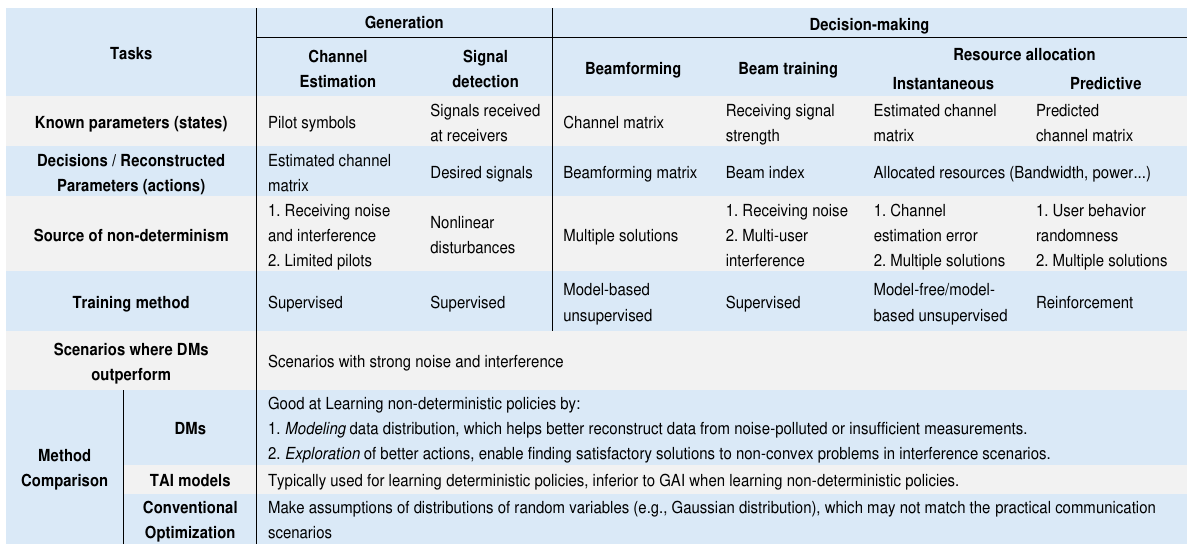}
	\caption{Applications of DMs for generation and decision-making tasks.}
	\label{fig:determination}
\end{figure*}

\vspace{-1mm}
\subsection{Beamforming}
The state and action of the beamforming policy are respectively the estimated channel matrix and the beamforming matrix.
In the multi-user system, the global optimal beamforming matrix is usually unavailable due to the non-convexity of optimization problems in interference scenarios. Hence, \emph{model-based unsupervised learning} can be exploited to train a DM for learning the policy.

Due to the non-convexity of the beamforming optimization problem in interference scenarios, the same state may correspond to multiple optimal actions that differ significantly. By starting with noise and adding noise on each reverse step, the DM can jump out of local optimal solutions to explore better actions.

\emph{Case study}: 
We use a case study to evaluate the advantage of DMs for exploring better actions. Consider learning digital beamforming at the base station (BS) for transmitting to four single-antenna users, aiming at maximizing the SE while satisfying a total power constraint. The channel coefficients from the BS to the users are assumed to be accurately estimated. The optimal solution structure in \cite{Precod_Opt_Structure} is leveraged such that only the power allocation for the users needs to be learned. Then, the beamforming matrix can be recovered with the solution structure. We compare the learning performance of the following four architectures, 
\begin{itemize}
	\item \textbf{DM (FNN) and DM (GNN)}: These are two DMs each with the noise-prediction DNN being a fully-connected neural network (FNN) and a graph neural network (GNN), respectively. The GNN can leverage the permutation property of the beamforming policy, i.e., the policy is not affected by changing the orders of users and antennas at the BS.
	\item \textbf{FNN and GNN}: These are respectively a standalone FNN and GNN for learning the mapping from the channel matrix to the power allocated to the users.
\end{itemize}

\begin{figure}
	\centering
	\includegraphics[width=.95\linewidth]{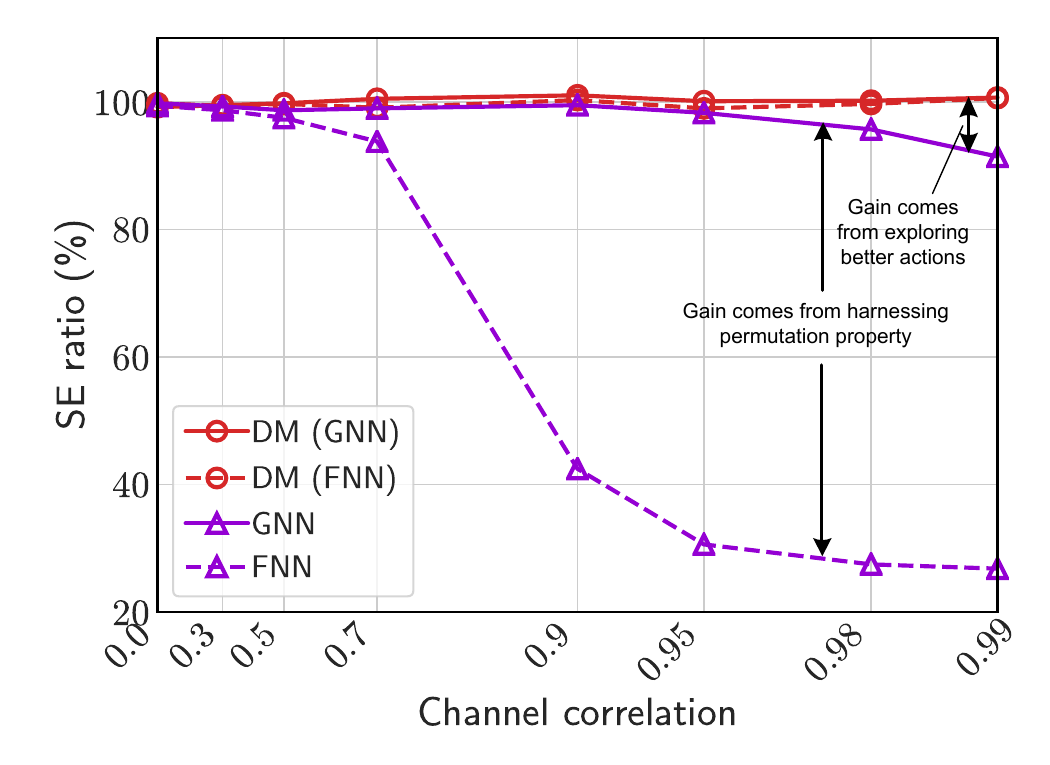}
	\caption{Performance of different architectures versus channel correlation. The numbers of BS antennas and users are respectively eight and four. The SNR is 10 dB.}
	\label{fig:performance}
\end{figure}

In Fig. \ref{fig:performance}, we show the learning performance of these architectures versus the correlation between user channel vectors. The performance is measured by the ratio of SE achieved by each architecture to the SE achieved by the weighted-minimum-mean-squared-error algorithm (called SE ratio). The channel vector of each user is generated as the weighted summation of two terms, which follow the Rayleigh distribution and are respectively the same and different among users. The channel correlation can be controlled with the weighting factor.

We can see from the results that the DM (GNN) and DM (FNN) respectively perform better than GNN and FNN, especially when the channel correlation is strong. The reason is as follows. Consider an extreme case that the channel vectors of all the users are almost the same. To avoid strong inter-user interference that degrades SE, the action (i.e., beamforming matrix) is optimal at four peaks. In each peak, only one user is allocated with full power while the other three users are allocated with zero power. The FNN and GNN tend to learn a single-modal policy with an output of equal power allocation due to identical user channels. By contrast, the DMs can explore better actions, as explained above. Furthermore, the GNN performs better than the FNN in strong channel correlation regimes, because the permutation property of the beamforming policy is incorporated with the GNN.

\vspace{-1mm}
\subsection{Beam Training}
Beam training involves sweeping candidate beamforming vectors (codewords) in a beamforming codebook to select the best codeword for transmission, such as the one maximizing the received power.
The state of the beam training policy is the receiving signal power by transmitting with the codewords (or a subset of codewords) in the codebook. The action is the selecting decision.
A DM can be trained \emph{in a supervised manner} to learn the policy. 

Polluted by receiving noise and multi-user interference, the strongest received signal may not always correspond to the best codeword, such that the beam training policy is non-deterministic. By iteratively denoising through learned reverse processes, DMs can effectively mitigate the impact of strong noises and interference by approximating their unknown and complex distributions \cite{GAI-signal-detection-TCOM2021}, such that the reliability and accuracy of beam training can be enhanced.

\vspace{-1mm}
\subsection{Resource Allocation}
The resources in the multiple antenna system, including power, bandwidth and time, can be optimized to improve system performance such as SE and energy efficiency. The resource allocation can either be optimized instantaneously or in a long term. 
\begin{itemize}
	\item \textbf{Instantaneous optimization}: The state and action can be the estimated channel matrix and the allocated resource. A DM can be trained in \emph{a model-free or a model-based unsupervised manner} to learn the resource allocation policy.
	\item \textbf{Long-term optimization}: This involves predicting future channel matrices over multiple time slots and optimizing resource allocation accordingly. The long-term resource allocation (also called predictive resource allocation) can be modeled as a Markov decision process. \emph{Deep reinforcement learning} can be leveraged to train a DM for maximizing long-term rewards.
\end{itemize}

Non-determinism in resource allocation arises from estimation errors (for instantaneous policies) or channel prediction inaccuracies due to randomness of user behaviors (for long-term policies). As mentioned above, DMs can better determine the actions than the TAI models by modeling the distributions of channels and user behaviors. Again, due to the presence of interference, the optimization problems are non-convex, such that multiple optimal actions may correspond to the same state. DMs are powerful for exploring better actions instead of struggling in a less optimal one.

Fig. \ref{fig:determination} summarizes the known parameters (states) and decisions/reconstructed parameters (actions) for each task, as well as the advantages of DMs.

\vspace{-1mm}
\section{Application of DMs for Emerging Multiple Antenna Techniques}
\begin{figure*}
	\centering
	\includegraphics[width=\linewidth]{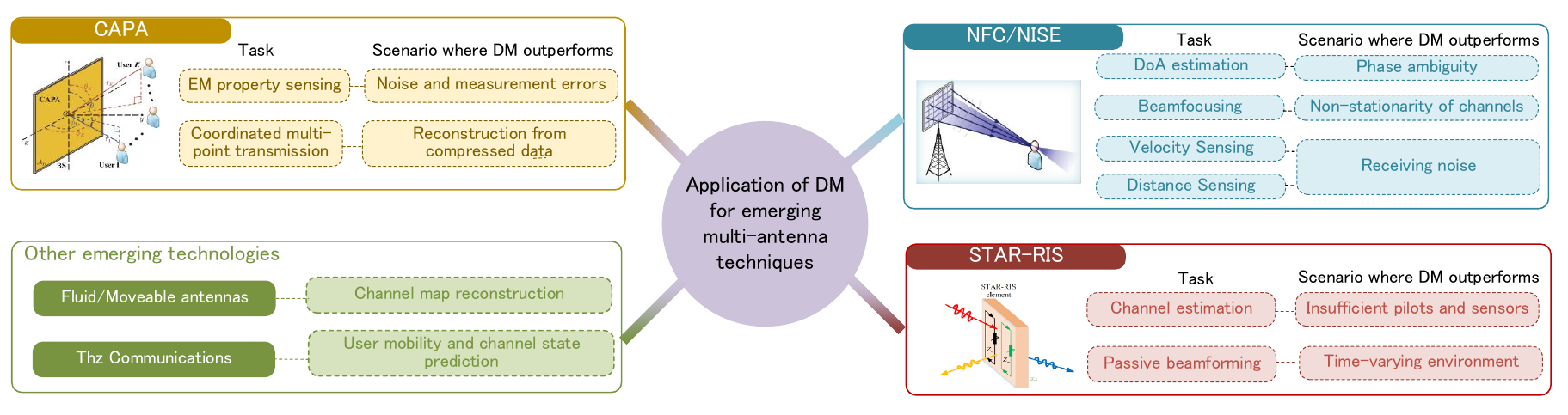}
	\caption{Applications of DMs for emerging multiple antenna technologies.}
	\label{fig:multiple antenna}
\end{figure*}

This section explores how DMs can be applied to emerging multiple antenna technologies, including near-field communications/sensing (NFC/NISE), STAR-RIS and CAPA. Fig. \ref{fig:multiple antenna} summarizes the associated tasks.

\vspace{-1mm}
\subsection{NFC/NISE}
In 6G networks employing extremely large aperture arrays and operating at exceedingly high frequencies, the near-field region extends to the scale of hundreds of meters. Unlike far-field propagation, electromagnetic (EM) waves in the near-field exhibit unique properties.
The non-negligible near-field region necessitates the investigation of NFC and NISE. DMs can be exploited in the NFC/NISE technologies focusing on the following areas,
\begin{itemize}
	\item \textbf{DoA estimation}: When the antenna spacing exceeds half-wavelength, the DoA is not uniquely determined by the phase difference between the signals received by adjacent antennas \cite{GAI-ISAC-WCM2024}. DMs can be trained in a supervised manner to learn the distribution of DoA for more accurate estimation.
	\item \textbf{Beamfocusing}: Near-field channels experience uneven path loss across the aperture and are prone to blockage by obstacles, posing challenges to accurate beamfocusing. Meanwhile, channels in high-frequency bands exhibit strong structures characterized by limited parameters, such as DoA. DMs can capture the channel structure for accurate beamfocusing. 
	\item \textbf{Velocity/distance sensing}: The measurements can be ambiguous due to strong receiving noise. 
	DMs enhance sensing accuracy by leveraging noisy observations and iteratively refining them into more precise estimates.
\end{itemize}

\vspace{-1mm}
\subsection{STAR-RIS}
As a candidate technology in 6G, simultaneously transmitting and reflecting reconfigurable intelligent surface (STAR-RIS) enables both transmission and reflection of signals incident on the programmable metasurface. STAR-RIS has the potential to smartly control the full-space propagation environment, thus enhancing signal strength, reducing interference, and improving energy efficiency. DMs can be applied in the STAR-RIS technologies in the following two aspects:
\begin{itemize}
	\item \textbf{Channel estimation}: 
	Owing to the passive nature, acquiring full-space transmitting/reflecting channels of STAR-RIS relies on extra active sensors, which enhances estimation accuracy but increases hardware costs. 
	This constitutes a research interest in generating STAR-RIS channel parameters by DMs exploiting fewer sensors and pilot measurements without sacrificing performance.
	\item \textbf{Passive Beamforming}: Coordination between transmission and reflection of signals is required for optimizing system performance under users' mobility. 
	DM-enabled generative AI algorithm can be explored to capture the distribution of the dynamic environment to better balance the reflection and transmission coefficients for passive beamforming, thus improving users' communication qualities without acquiring perfect channel information. 
\end{itemize}

\vspace{-1mm}
\subsection{CAPA} 
The continuous aperture array (CAPA) can be regarded as a conventional spatially discrete array (SPD) comprising an infinite number of antennas with infinitesimal spacing, thereby providing a high spatial degree-of-freedom and channel capacity. In CAPA-aided systems, DMs can be leveraged in the following tasks,
\begin{itemize}
	\item\textbf{EM property sensing}: CAPA offers higher spatial resolutions, enabling more accurate EM property sensing. TAI-enabled sensing methods are often sensitive to noise and measurement errors, which yields unreliable results. DMs excel at capturing the distribution of point clouds of a target, such that the EM property can be better constructed from the estimated sensing channels \cite{DM-ISAC}.
	\item\textbf{Coordinated multi-point transmission}: Multiple BSs equipped with CAPAs can be coordinated to serve users in a larger area. To reduce the overhead of information exchange between BSs, data (e.g., channel state information, transmission data, and user positions) need to be compressed. DMs can be beneficial by learning the data distribution and enabling accurate reconstruction from compressed data, such that multiple BSs can be coordinated with shared information for optimizing overall system performance.
\end{itemize}

\subsection{Other Emerging Multiple Antenna Technologies}

DMs can also be applied to other emerging multiple antenna technologies. Taking fluid/movable antenna technology as an example, DMs can be leveraged to reconstruct channel maps from sparse measurements with low training overhead, which further enables antenna position optimization. In Terahertz (Thz) communications, DMs excel at predicting channel states by learning the distribution of user mobility patterns, enabling proactive decision-making such as predictive handovers.

\vspace{-1mm}
\section{Conclusions and Promising Research Directions}\label{sec:future-works}
In this article, the potential applications of DMs to multiple antenna technologies have been unveiled. To provide a general guideline, a unified framework of applying DMs for policy learning have been put forward. Then, the applications of DMs in decision-making tasks and generation tasks have been explored, and the reasons why DMs are suitable for learning these non-deterministic policies were analyzed. The numerical results of a case study of learning beamforming with a DM have also been provided.  
There are still numerous open research directions in this area. Some of them are listed as follows.
\begin{itemize}
	\item \textbf{Improve learning efficiency of DMs}: High-dimensional state and action spaces in large-scale networks pose significant challenges on the training overhead for DMs.  
	To reduce the overhead, domain knowledge can be incorporated with the DMs. For example, the noise-prediction DNN can be designed as a GNN that can leverage the permutation property of a policy to be learned \cite{LSJ_MultiDim_GNN_2022}. The DNN can also be designed as a model-based structure such as a deep unfolding network to exploit the knowledge of the iterative procedure of an algorithm.
	\item \textbf{Theoretical understanding of DMs}: 
	Research on DMs for wireless communications is still in its early stages, with limited theoretical analysis to explain their effectiveness or guarantee performance. Leveraging mathematical models, such as optimization problems and numerical algorithms abundant in wireless tasks, can bridge this gap. For instance, the step-by-step architecture of DMs resembles iterative numerical algorithms, where random initialization parallels reverse diffusion from Gaussian noise. Such analogies can offer valuable insights to advance theoretical understanding.
	\item \textbf{Lightweight deployment of DMs}: 
	DMs usually suffer from high computational complexity during the inference phase, posing challenges for energy-efficient and real-time applications in large-scale multiple antenna communications. Research directions include model compression (e.g., pruning, quantization, and knowledge distillation), hardware acceleration (e.g., FPGA), and hybrid frameworks integrating diffusion models with simpler approximations.
\end{itemize}

\vspace{-1mm}

\bibliography{IEEEabrv,GJ}

\begin{thebibliography}{10}
\providecommand{\url}[1]{#1}
\csname url@samestyle\endcsname
\providecommand{\newblock}{\relax}
\providecommand{\bibinfo}[2]{#2}
\providecommand{\BIBentrySTDinterwordspacing}{\spaceskip=0pt\relax}
\providecommand{\BIBentryALTinterwordstretchfactor}{4}
\providecommand{\BIBentryALTinterwordspacing}{\spaceskip=\fontdimen2\font plus
\BIBentryALTinterwordstretchfactor\fontdimen3\font minus
  \fontdimen4\font\relax}
\providecommand{\BIBforeignlanguage}[2]{{%
\expandafter\ifx\csname l@#1\endcsname\relax
\typeout{** WARNING: IEEEtran.bst: No hyphenation pattern has been}%
\typeout{** loaded for the language `#1'. Using the pattern for}%
\typeout{** the default language instead.}%
\else
\language=\csname l@#1\endcsname
\fi
#2}}
\providecommand{\BIBdecl}{\relax}
\BIBdecl

\bibitem{NFC-survey}
Y.~Liu, C.~Ouyang, Z.~Wang, J.~Xu, X.~Mu, and A.~L. Swindlehurst, ``Near-field
  communications: A comprehensive survey,'' \emph{IEEE Commun. Surveys Tuts.},
  2024, Early access.

\bibitem{Survey-DM-DRL-NetwOpt-2024}
H.~Du, R.~Zhang, Y.~Liu \emph{et~al.}, ``Enhancing deep reinforcement learning:
  A tutorial on generative diffusion models in network optimization,''
  \emph{IEEE Commun. Surveys Tuts.}, vol.~26, no.~4, pp. 2611--2646, 2024.

\bibitem{GvsD}
E.~Argouarc'h, F.~Desbouvries, E.~Barat \emph{et~al.}, ``Generative vs.
  discriminative modeling under the lens of uncertainty quantification,''
  \emph{arXiv:2406.09172}, 2024.

\bibitem{DM-DRL}
L.~Yang, Z.~Huang, F.~Lei, Y.~Zhong, Y.~Yang, C.~Fang, S.~Wen, B.~Zhou, and
  Z.~Lin, ``Policy representation via diffusion probability model for
  reinforcement learning,'' \emph{arXiv:2305.13122}, 2023.

\bibitem{XXX}
X.~Xu, X.~Mu, Y.~Liu \emph{et~al.}, ``Generative artificial intelligence for
  mobile communications: A diffusion model perspective,'' \emph{IEEE Commun.
  Mag.}, 2024, Early access.

\bibitem{GAI-signal-detection-TCOM2021}
K.~He, L.~He, L.~Fan \emph{et~al.}, ``Learning-based signal detection for
  {MIMO} systems with unknown noise statistics,'' \emph{IEEE Trans. Commun.},
  vol.~69, no.~5, pp. 3025--3038, 2021.

\bibitem{GAI-ISAC-WCM2024}
J.~Wang, H.~Du, D.~Niyato \emph{et~al.}, ``Generative {AI} for integrated
  sensing and communication: Insights from the physical layer perspective,''
  \emph{IEEE Wireless Commun.}, vol.~31, no.~5, pp. 246--255, Oct. 2024.

\bibitem{DDPM}
J.~Ho, A.~Jain, and P.~Abbeel, ``Denoising diffusion probabilistic models,''
  \emph{NeuIPS}, 2020.

\bibitem{Eisen_Unspv_TSP2019}
M.~Eisen, C.~Zhang, L.~F.~O. Chamon \emph{et~al.}, ``Learning optimal resource
  allocations in wireless systems,'' \emph{IEEE Trans. Signal Process.},
  vol.~67, no.~10, pp. 2775--2790, May 2019.

\bibitem{SCJ-modelfree}
D.~Liu, C.~Sun, C.~Yang \emph{et~al.}, ``Optimizing wireless systems using
  unsupervised and reinforced-unsupervised deep learning,'' \emph{IEEE Netw.},
  vol.~34, no.~4, pp. 270--277, Feb. 2020.

\bibitem{Precod_Opt_Structure}
E.~Bj{\"o}rnson, M.~Bengtsson, and B.~Ottersten, ``Optimal multiuser transmit
  beamforming: A difficult problem with a simple solution structure,''
  \emph{IEEE Signal Process. Mag.}, vol.~31, no.~4, pp. 142--148, July 2014.

\bibitem{DM-ISAC}
Y.~Jiang, F.~Gao, S.~Jin, and T.~J. Cui, ``Electromagnetic property sensing
  based on diffusion model in {ISAC} system,'' \emph{IEEE Trans. Wireless
  Commun.}, 2024, Early access.

\bibitem{LSJ_MultiDim_GNN_2022}
S.~Liu, J.~Guo, and C.~Yang, ``Multidimensional graph neural networks for
  wireless communications,'' \emph{IEEE Trans. Wireless Commun.}, vol.~23,
  no.~4, pp. 3057--3073, April 2024.

\end{thebibliography}

\begin{IEEEbiographynophoto}{Jia Guo} (Member, IEEE) is currently a Postdoctoral Researcher with Queen Mary University of London, U.K.
	He received his B.S. degree in electronics engineering, and Ph.D. degree in information and communication engineering from Beihang University, China, in 2016 and 2023, respectively. His research interests include AI for wireless communications.	
\end{IEEEbiographynophoto}

\vspace{-3mm}

\begin{IEEEbiographynophoto}{Xiaoxia Xu} (Member, IEEE) is currently a Postdoctoral Researcher at Queen Mary University of London, U.K. She received her B.Eng. and Ph.D. degree from Wuhan University in 2017 and 2023, respectively. She was also a visiting student with the Queen Mary University of London from 2021 to 2022. Her research interests include NOMA, mmWave/THz MIMO, AI for wireless
	networks, and mobile edge generation.
\end{IEEEbiographynophoto}

\vspace{-3mm}

\begin{IEEEbiographynophoto}{Yuanwei Liu} (Fellow, IEEE) is a (tenured) Full Professor at The University of Hong Kong and a visiting professor at Queen Mary University of London. His research interests include NOMA, RIS/STARS, Integrated Sensing and Communications, Near-Field Communications, and machine learning. He serves as a Co-Editor-in-Chief of IEEE ComSoc TC Newsletter, an Area Editor of IEEE TCOM and CL, and an Editor of the IEEE COMST/TWC/TVT/TNSE.
\end{IEEEbiographynophoto}

\vspace{-3mm}

\begin{IEEEbiographynophoto}{Arumugam Nallanathan} (Fellow, IEEE) (a.nallanathan@qmul.ac.uk) is a professor and the head of the Communication Systems Research (CSR) group in Queen Mary University of London. His research interests include beyond 5G wireless networks, the Internet of Things, and AI for Wireless Communications.
\end{IEEEbiographynophoto}

\end{document}